\documentclass[final,3p,times,numbers,sort&compress]{elsarticle}   
\usepackage{amsmath,amssymb,stmaryrd,amstext,amsthm}
\usepackage{mathtools}
\usepackage{bm}	
\usepackage{caption}
\usepackage[hidelinks]{hyperref}
\usepackage{subcaption}
\usepackage{siunitx}
\usepackage{booktabs}
\usepackage[english]{cleveref}
\usepackage{lineno}
\usepackage{float}
\usepackage{graphicx}
\usepackage{verbatim}   % for math
\newcommand{\volsym}{\rlap{\kern.08em--}V} % Volume symbol

\usepackage{verbatim}

\def\tsc#1{\csdef{#1}{\textsc{\lowercase{#1}}\xspace}}
\tsc{WGM}
\tsc{QE}
\tsc{EP}
\tsc{PMS}
\tsc{BEC}
\tsc{DE}
\journal{} 

\begin{document}

\begin{frontmatter}



%\title{An improved formulation for the wake-added turbulence for intra-farm and farm-to-farm wake modeling}
\title{Wind-farm power prediction using a turbulence-optimized Gaussian wake model}
\author{Navid Zehtabiyan-Rezaie}
%\ead {zehtabiyan@mpe.au.dk}
\author{Josephine Perto Justsen}
%\ead {202010633@post.au.dk}
\author{Mahdi Abkar\corref{cor1}}
\cortext[cor1]{Corresponding author}
\ead {abkar@mpe.au.dk}

\address{Department of Mechanical and Production Engineering, Aarhus University, 8200 Aarhus N, Denmark}

\begin{abstract}
In this study, we present an improved formulation for the wake-added turbulence to enhance the accuracy of intra-farm and farm-to-farm wake modeling through analytical frameworks. Our goal is to address the tendency of a commonly used formulation to overestimate turbulence intensity within wind farms and to overcome its limitations in predicting the streamwise evolution of turbulence intensity beyond them. To this end, we utilize high-fidelity data and adopt an optimization technique to derive a refined functional form of the wake-added turbulence. We then integrate the achieved formulation with a widely used Gaussian wake model to study various intra-farm and farm-to-farm scenarios. The outcomes reveal that the new methodology effectively addresses the overestimation of power in both standalone wind farms and those impacted by upstream counterparts. Our new approach meets the need for accurate and lightweight models, ensuring the effective coexistence of wind farms within clusters as the wind-energy capacity rapidly expands.
\end{abstract}

%\begin{highlights} %%Research highlights
%\item An improved formulation for the wake-added turbulence is presented.
%\item The new formulation is integrated into a widely used Gaussian wake model. 
%\item Diverse intra-farm and farm-to-farm scenarios are analyzed.
%\item Enhanced power prediction for both isolated and downstream-influenced wind farms is achieved.   


\end{frontmatter}
%\linenumbers %% add line number


%\linenumbers

\section{Introduction} \label{Sec:Introduction}
Wind energy is crucial in transitioning the power-production sector towards cleaner and more sustainable solutions, supporting efforts to reduce greenhouse gas emissions and mitigate global warming. Driven by its benefits, many countries have set ambitious deployment targets for wind energy in the coming decades \cite{veers2023grand,GLOBALWINDREPORT2024}. Such a rapid expansion brings forth two significant challenges. The first challenge is the classic problem of quantification and management of \emph{intra-farm} wake effects. Wind farms harvest energy through fully coupled and multi-scale interactions with the atmosphere, creating regions characterized by high turbulence and reduced velocity behind the turbines, known as wakes. Due to short spacing between turbines, wakes do not have sufficient opportunity to fully recover through the turbulent-mixing mechanism. This translates into downstream turbines experiencing decreased power production and increased mechanical loads (see the review of Refs. \cite{Vermeer2003, Stevens2017, PorteAgel2019Review} and references therein). Effective intra-farm wake management through design, optimization, and control is an objective vigorously pursued to enhance the performance of wind farms. Given the rising trends in wind energy, particularly the increase in power density, it becomes even more imperative to quantify and mitigate these adverse effects \cite{HerbertAcero2014,Meyers2022,ABKAR2023_TAML}.

The second challenge pertains to the proximity of wind farms to one another, extending the concern to \emph{farm-to-farm} wake effects \cite{Platis2018}. As new wind farms are established, e.g., in favorable sites like the North Sea \cite{northSea2022}, they are frequently located near existing installations due to limited space. Field measurements reveal that the wake from a wind farm can stretch for tens of kilometers downstream \cite{Christiansen2005,Pettas2021,Pryor2024} and, thus, quantifying the farm-to-farm wake losses is crucial to ensure an effective coexistence within wind-farm clusters.

Advances in computational and numerical techniques have profoundly impacted the realm of wind-turbine and wind-farm wake modeling. Computational fluid dynamics (CFD) methods, such as large-eddy simulations (LES) and Reynolds-averaged Navier-Stokes (RANS) models, have been employed to explore wake-flow characteristics within an individual wind farm (see, for example, Refs. \cite{Wu2015,Stevens2018,RANS_extended, zehtab2024secondary}, among several notable works) and beyond/between wind farms (e.g., Refs. \cite{Eriksson2014,Rafalimanana2015,meng2021study,dong2022far,Stieren2022}, among others). While these techniques enable detailed investigations of the complex wake flows, they are not computationally efficient for practical applications that require fast-paced predictions \cite{Archer2018,ABKAR2023_TAML}. 
Therefore, achieving these goals critically relies on analytical wake-modeling frameworks. 

At its core, an analytical wake-modeling framework is grounded in a model designed to represent the wake behind an individual wind turbine \cite{jensen1983a, Frandsen2006, Bastankhah2014,Xie2015}. This model can be constructed by applying fundamental conservation laws and incorporating assumptions such as a predefined velocity-deficit profile and a specific functional form for wake expansion.
Since wind turbines within wind farms frequently encounter multiple wakes from upstream turbines, analytical wake-modeling frameworks must account for the cumulative effects of these interacting wakes. This is achieved by applying stand-alone wake models to each turbine individually and using superposition principles\cite{lissaman1979energy, katic1986simple, voutsinas1990analysis, Niayifar2016}.
For example, the Park model applies the conservation of mass to a control volume downwind of a turbine and assumes a top-hat shape for the velocity deficit \cite{jensen1983a}, integrated with an energy-deficit superposition technique \cite{katic1986simple}. It assumes a linear wake growth behind turbines which can be linked to the ambient turbulence level. Another widely used tool is the framework proposed by Niayifar and Port{\'{e}}-Agel \cite{Niayifar2016}, which builds upon the Gaussian single-turbine wake model of Bastankhah and Port{\'{e}}-Agel \cite{Bastankhah2014} and employs a linear superposition method to aggregate the impacts of overlapping wakes. It adopts a linear function for the streamwise variation of wake width behind a turbine, which is, in turn, a function of the incoming streamwise turbulence intensity. Analytical wake-modeling frameworks have seen extensive application in the literature (see, e.g., Refs. \cite{annoni2018,Portagel2018,simley2020,zehtab2022_2,Souaiby2024}, among others) and are integrated into tools such as FLORIS \cite{FLORIS} and PyWake \cite{pywake2023}.

One of the key features of recent wake models (e.g., Refs. \cite{Niayifar2016,Bastankhah2021,HVolker2022,Bastankhah2024}) is the connection between the wake recovery rate behind turbines and the incoming turbulence intensity, while also considering the turbulence generated by the wake.
Various empirical correlations exist in the literature to parameterize the wake-added turbulence (e.g., Refs. \cite{quarton1990, Crespo1996, frandsen2007turbulence, Xie2015, Ishihara2018,li2022novel}). One widely adopted method, introduced by Crespo and Hern\'andez \cite{Crespo1996}, predicts the maximum added turbulence, rather than the disk-averaged value, occurring near the top-tip and side-tip regions downwind of a turbine, which are locations associated with strong shear \cite{abkar2015influence}. This often leads to an overestimation of the wake growth rate behind individual turbines and, consequently, an inaccurate projection of the power produced by the farm \cite{zehtab2023_note}. Additionally, the correlation of Crespo and Hern\'andez \cite{Crespo1996} is validated within a downstream distance of $5 < x/d_0 < 15$, raising questions about whether it can be used in conjunction with a wake model to study farm-to-farm wake effects.

In this paper, we aim to achieve two primary objectives: 1) systematically investigating the validity of using the empirical correlation proposed by Crespo and Hern\'andez \cite{Crespo1996} to represent the evolution of turbulence intensity within and beyond wind farms, and 2) enhancing the wake-added turbulence formulation for improved intra-farm and farm-to-farm wake modeling. To this end, we employ a grid-search optimization methodology to refine the correlation based on the LES data of a single wind turbine. We then utilize two suites of in-house LESs to thoroughly investigate the evolution of streamwise turbulence intensity inside the farms and their downstream. Subsequently, we use different LES cases to assess the adequacy of the Niayifar and  Port{\'{e}}-Agel \cite{Niayifar2016} model, coupled with the original and optimized wake-added turbulence formulations, applied for intra-farm and farm-to-farm wake modeling. 
The rest of the manuscript is structured as follows: Section~\ref{Sec:Method} details the adopted methodology, and Section~\ref{Sec:Results} presents the results of the intra-farm and farm-to-farm wake investigations compared against LES predictions. Finally, the main conclusions from this work are summarized in Section~\ref{Sec:Conclusions}.

\section{Methodology} \label{Sec:Method}
In this section, we initially introduce our analytical wake-modeling framework. We then focus on the formulations to incorporate wake-added turbulence within the wake model. Following these steps, we detail the high-fidelity LES datasets employed at various stages of this study. 

\subsection{Analytical wake-modeling framework} 
The analytical wake-modeling framework of Niayifar and Port{\'{e}}-Agel \cite{Niayifar2016} adopts an axisymmetric self-similar Gaussian distribution for the normalized velocity-deficit profiles at any position $(x,y,z)$ in the far-wake region behind turbine $j$ as
\begin{align}
    \frac{\Delta{u}_j(x,y,z)}{ {u}_{j,\text{in}} } &= \left(1 - \sqrt{1 - \frac{C_{T}d_0^2}{8\sigma^2} }\right) \exp \left(-\frac{y^2+z^2}{2\sigma^2} \right), \label{eq:calcDeltaU}
\end{align}

\noindent where $x$, $y$, and $z$ are coordinates with respect to the center of turbine $j$ in the streamwise, spanwise, and vertical directions, respectively. $\Delta{u}_j$ is the velocity deficit in the wake defined with respect to ${u}_{j,\text{in}}$ which is the rotor-averaged incoming-flow velocity of turbine $j$. $C_T$ denotes the turbine's thrust coefficient, and $d_0$ is the turbine's rotor diameter.
The standard deviation of the Gaussian velocity deficit ($\sigma$) varies linearly with distance, following the relation $\sigma/d_0 = k^*_j x /d_0 + \varepsilon$, in which $\varepsilon$ is defined as $0.2 \sqrt{\beta}$ with $\beta = 0.5\left(1+\sqrt{1-C_T}\right)/\sqrt{1-C_T}$. The parameter $k^*_j$, known as the wake growth rate behind turbine $j$, is defined as
\begin{align}
    k^*_j=0.38 I_{j,\text{in}} + 0.004, \label{eq:kstar}
\end{align}
\noindent where $I_{j,\text{in}}$ is the incoming streamwise turbulence intensity at the $j$th turbine's location in the absence of the rotor \cite{Niayifar2016}. 

Here, we utilize a linear superposition of velocity deficit suggested by Niayifar and Port{\'{e}}-Agel \cite{Niayifar2016} as
\begin{align}
    {u}(x,y,z) = {u}_\infty - \displaystyle\sum_{j} \Delta{u}_j, 
    \label{eq:supVel}
\end{align}
\noindent where ${u}_\infty$ is the undisturbed inflow velocity for the wind farm. Hereafter, we will refer to this analytical wake-modeling framework as the Gaussian model (GM). 

\subsection{Wake-added turbulence formulations} 
The approach proposed by Niayifar and Port{\'{e}}-Agel \cite{Niayifar2016} evaluates the wake growth rate behind each turbine (through Eq.~(\ref{eq:kstar})) based on the value of the incoming streamwise turbulence intensity. The evolution of the streamwise turbulence intensity within the farm is incorporated into the model as
\begin{align}
    I_{\text{w}} = \sqrt{I_\infty^2+I_+^2}, 
    \label{eq:I_wake}
\end{align}
\noindent where $I_{\text{w}}$ is the streamwise turbulence intensity in the wake, $I_\infty$ is the streamwise turbulence intensity in the free stream, and $I_{+}$ is the wake-added turbulence intensity \cite{Crespo1996}. 

As our baseline wake-added turbulence formulation, we choose the widely adopted method introduced by Crespo and Hern\'andez \cite{Crespo1996} and, hereafter, refer to it as $I_{+,\text{org.}}$. According to this formulation, the maximum added turbulence intensity in the region expanding with range $5 < x/d_0 < 15$ downstream a turbine is estimated as
\begin{align}
     I_{+,\text{org.}} = 0.73 a^{0.8325} I_\infty^{-0.0325} \left({x}/d_0\right)^{-0.32},
    \label{eq:I_added_org}
\end{align}
\noindent where $a = (1 - \sqrt{1 - C_T})/{2} $ is the axial induction factor, and $x$ denotes the streamwise location behind the turbine. Focusing on the functional form of this relation, one can see that it inversely correlates with the ambient turbulence level. However, some confusion propagates in the literature regarding the correct application of this correlation, particularly concerning the negative exponent on $I_\infty$, as emphasized in our previous work \cite{zehtab2023_note}. 

The correlation proposed by  Crespo and Hern\'andez \cite{Crespo1996} outputs the maximum wake-added turbulence value. Hence, it tends to overestimate incoming turbulence levels and, consequently, the wake growth rates\cite{zehtab2023_note}. Additionally, the model's formulation regarding the streamwise location requires refinement, as its initial validation is based on data collected in the range $5 < x/d_0 < 15$ behind a turbine.
To simultaneously correct the magnitude of wake-added turbulence and improve the accuracy of its streamwise evolution (for $x \gg 15d_0$), while preserving its exponents on the induction factor and ambient turbulence level, we adopt a new coefficient $C_1$ and a new exponent $C_2$. We will optimize $C_1$ and $C_2$ in this study to achieve a modified formulation for the added turbulence as
\begin{align}
     I_{+,\text{mod.}} =\textcolor{black}{C_1} a^{0.8325} I_\infty^{-0.0325} \left({x}/d_0\right)^{\textcolor{black}{C_2}}.
    \label{eq:I_added_new}
\end{align}

We need a strategy to consider the impact of overlapping wakes on the turbulence intensity within the farm. There exist two approaches in this context, either considering the cumulative effect of all upstream turbines or just the impact of the closest upstream turbine \cite{li2023novel}. Niayifar and Port{\'{e}}-Agel \cite{Niayifar2016} chose the second method based on the observations made in prior studies, e.g., Ref. \cite{frandsen1999integrated}. Here, we will also adopt a similar non-cumulative strategy that can be formulated as
\begin{align}
    I_{+_j} = \max \left(\frac{A_\text{w}}{A_0} I_{+_{kj}} \right) , 
    \label{eq:supTI}
\end{align}
\noindent where $A_\text{w}/A_0$ denotes the ratio of the cross-sectional area of the wake to the rotor's area, and $I_{+_{kj}}$ is the added streamwise turbulence intensity due to turbine $k$ at the location of turbine $j$. 

\subsection{Description of high-fidelity datasets} \label{sec:cases}
\begin{figure}[!ht]
	\centering
    \includegraphics[width=0.6\textwidth]{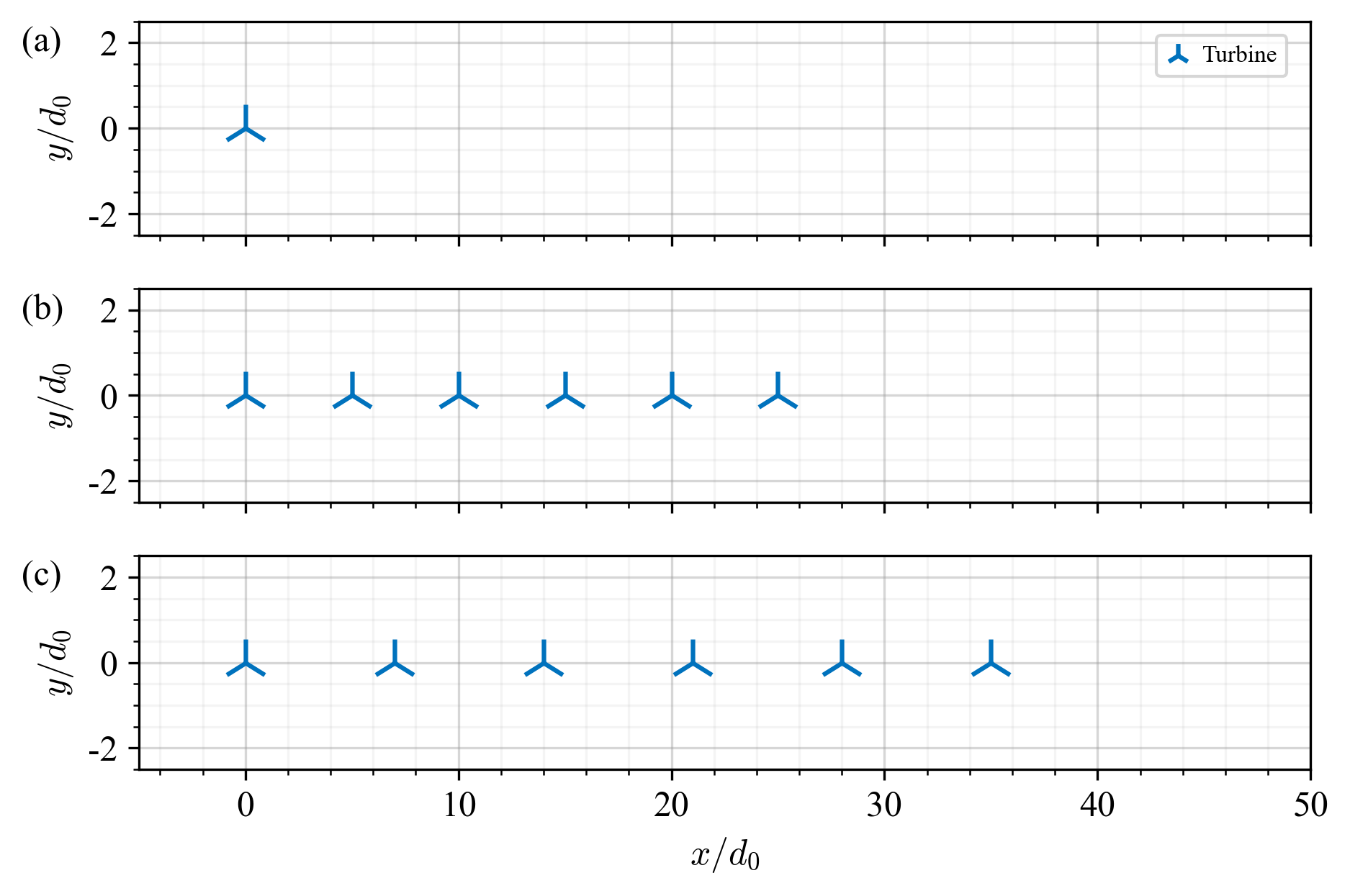}
	\caption{The layouts of the three wind-farm cases utilized in the parameter-optimization and model-validation stages: (a) Single-turbine, (b) 6 inline turbines with a streamwise spacing of $5d_0$, and (c) 6 inline turbines with a streamwise spacing of $7d_0$. The subplots correspond to Cases (Opt.), (V-1), and (V-2), from top to bottom.}
	\label{fig:Intra1_2}
\end{figure}

In this study, we leverage different sets of LES data to achieve our objectives introduced earlier. Table~\ref{tab:case_summary} holds detailed information on these datasets including their utilization stage, wind-farm layout, turbines' specifications and operating condition, and the undisturbed flow characteristics at the hub height.
Initially, we will use three in-house LES datasets \cite{Eidi2022} including a standalone wind turbine, an inline six-turbine array with a streamwise spacing of $5d_0$, and a six-turbine array with a spacing of $7d_0$ between the consecutive turbines. The corresponding layouts are given in Figures~\ref{fig:Intra1_2}(a) to (c). 
The evolution of streamwise turbulence intensity behind the standalone turbine (Case (Opt.)) obtained from LES will be used as a reference to determine the optimal coefficients in Eq.~(\ref{eq:I_added_new}). Afterward, we will evaluate the distribution of streamwise turbulence intensity downstream of Cases (V-1) and (V-2). Following that, we will validate the model by focusing on the streamwise turbulence intensity at the location of turbines in these cases.

After the optimization and validation stages, we will first focus on intra-farm wake modeling. For this purpose, two $12 \times 6$ arrays of wind turbines, one aligned (Case (IF-1)) and one staggered (Case (IF-2)), are selected with LES data presented in the study of Stieren and Stevens \cite{Stieren2022}. We then delve deeper into intra-farm wake modeling to study the model's prediction under different incoming wind directions by applying it to Horns Rev 1 wind farm (Case (IF-3)) and comparing the results against LESs performed by Port{\'{e}}-Agel \textit{et al.} \cite{PorteAgel2013}.

In the last part of our study, we focus on farm-to-farm wake modeling, studying the wake losses in a downstream wind farm due to the presence of an upstream counterpart, with the distancing of $S_\text{WF}$ varying from 5-15 km. To this end, five combinations will be investigated (named as Cases (FF-1) to (FF-5)) and the normalized power of the downstream wind farm will be compared against data from Ref. \cite{Stieren2022}.

\begin{table}[ht]
\centering
\caption{Characteristics of different cases studied for optimization and validation (V) of the new wake-added turbulence model, followed by intra-farm (IF) and farm-to-farm (FF) cases. Here, $S_x$ and $S_y$ denote the streamwise and spanwise turbine spacing, respectively, and $z_\text{h}$ is the turbine's hub height. $S_\text{WF}$ stands for the streamwise distance between the upstream and downstream wind farms.}
\label{tab:case_summary}  
\begin{tabular}{>{\arraybackslash}p{.8cm}>{\arraybackslash}p{9.2cm}*{7}{>{\centering\arraybackslash}p{0.7cm}}}
\hline
Case	       &	Layout	             &	$d_0$ [m] 	&	$z_\text{h}$ [m]   &	$C_T$   &	${u}_\infty$ [m/s]  & $I_\infty$ [\%]   \\  \hline  \hline 
Opt.	   &	Stand-alone turbine	     &		80 	&	70 	       &	0.75                      &	8.0 	    & 7.25        \\	
V-1	   & 6 inline turbines; $S_x=5d_0$		    	           &	80 	&	70 	       &	0.75	                      &	8.0	    & 7.25      	\\	
V-2    &	6 inline turbines; $S_x=7d_0$	                  &	80 	&	70 	       &	0.75	                      &	8.0     & 7.25    	\\	\hline  %\cdashline{1-7}  
IF-1	   &	$12 \times 6$ aligned array; $S_x=7d_0, S_y=5d_0$	        	           &	120 	&	100 	       &	0.75	                      &	9.5 	    & 7.03        	\\
IF-2	   &	$12 \times 6$ staggered array; $S_x=7d_0, S_y=5d_0$	    	           &	120 	&	100 		       &	0.75 &	9.5 	    & 7.03       \\ 
IF-3	   &	Horns Rev 1 wind farm	            &	80 	&	70 	       &	0.80	                  &	8.0	    & 9.63      \\	\hline  %\cdashline{1-7} 
FF-1	   &	Upst. and downst. farms: $12 \times 6$ align.; $S_\text{WF}=10 \text{ km}$	           &	120 	&	100 	       &	0.75	                  &	9.5 	    & 7.03      	\\
FF-2	   &	Upst. farm: $12 \times 6$ stag.; downst. farm: $12 \times 6$ align.; $S_\text{WF}=10 \text{ km}$		           &	120 	&	100 		       &	0.75	                  &	9.5 	    & 7.03      \\
FF-3	   &	Upst. and downst. farms: $12 \times 6$ stag.; $S_\text{WF}=5 \text{ km}  = 41.7d_0$            &	120	&	100		       &	0.75	                  &	9.5     & 7.03        \\
FF-4	   &	Upst. and downst. farms: $12 \times 6$ stag.; $S_\text{WF}=10 \text{ km} = 83.3d_0 $	     &	120 	&	100 	       &	0.75	                  &	9.5 	    & 7.03     \\
FF-5	   &	Upst and downst. farms: $12 \times 6$ stag.; $S_\text{WF}=15 \text{ km} = 125d_0$      &	120 	&	100        &	0.75	                  &	9.5     & 7.03   \\   \hline 
\end{tabular}
\end{table}

\begin{figure}[!ht]
	\centering
    \includegraphics[width=0.8\textwidth]{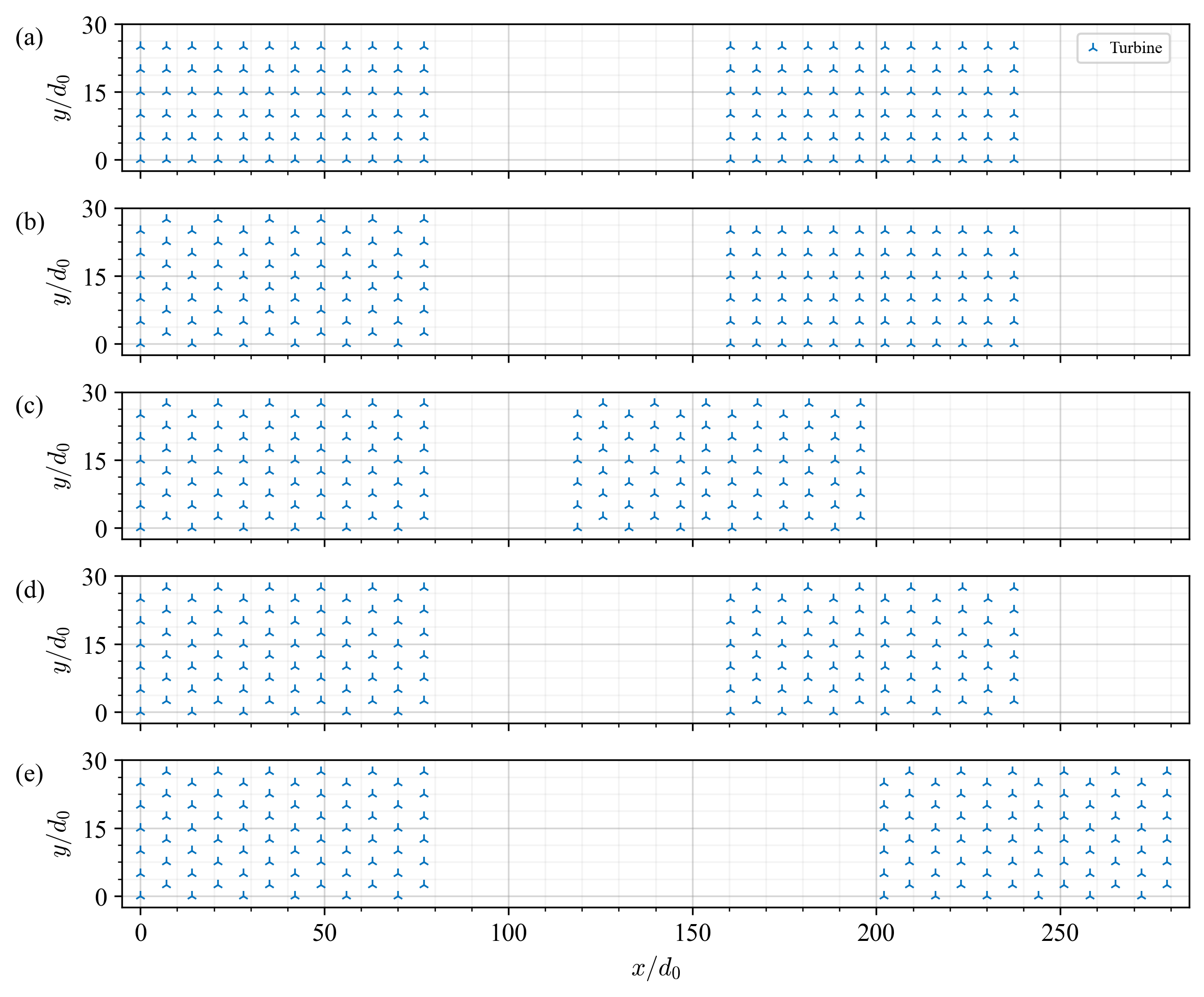}
	\caption{Illustration of the layouts of cases investigated for farm-to-farm wake modeling, i.e., Cases (FF-1) to (FF-5), depicted in subplots from top to bottom. Here, the spacing between the farms, denoted as $S_\text{WF}$, ranges from 5 to 15 km.}
	\label{fig:Inter1_5}
\end{figure}
\section{Results and discussion} \label{Sec:Results}

In this section, initially, we focus on the optimization procedure adopted to derive the new wake-added turbulence formulation. We then further validate the model by looking into the evolution of streamwise turbulence intensity within and beyond two distinct wind farms. Finally, we conduct a systematic performance assessment by first applying the GM integrated with both the original and optimized wake-added turbulence formulations to intra-farm cases and, subsequently, to farm-to-farm cases, followed by a detailed discussion of the results.

\subsection{Optimization and validation}
Eq.~(\ref{eq:I_added_new}) has two unknowns, i.e., $C_1$ and $C_2$, that need to be determined to construct the new wake-added turbulence formulation. To this end, we employ a grid-search optimization where these coefficients are found by minimizing the mean squared error between the evolution of the disk-averaged streamwise turbulence intensity behind a single turbine, predicted through the modified formulation and LES results. Following this procedure, the optimized wake-added turbulence formulation is found to be
\begin{align}
     I_{+,\text{opt.}} =\textcolor{black}{0.9} a^{0.8325} I_\infty^{-0.0325} \left({x}/d_0\right)^{\textcolor{black}{-0.56}}.
    \label{eq:I_added_opt}
\end{align}

\begin{figure}[!ht]
	\centering
    \includegraphics[width=0.6\textwidth]{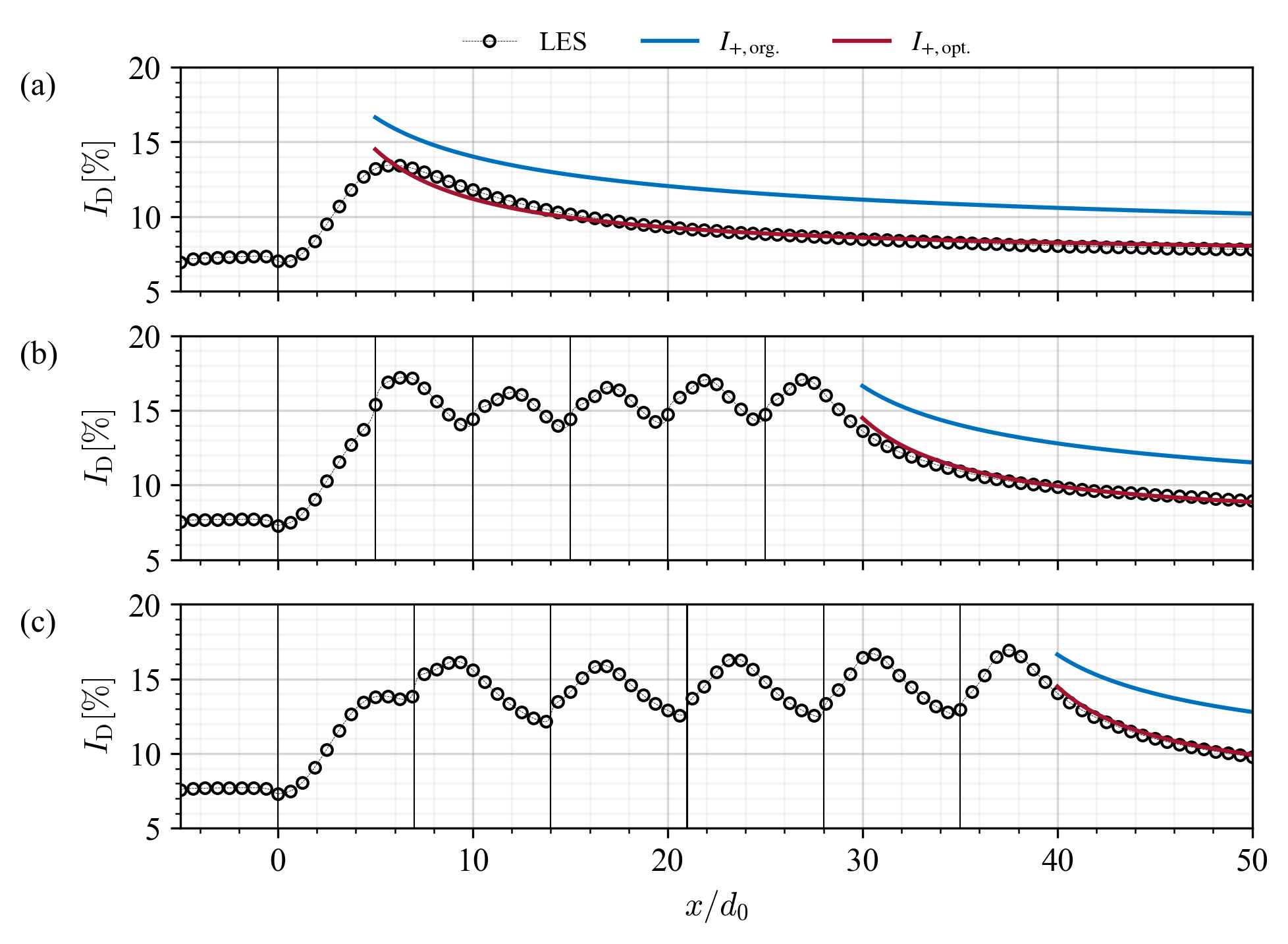}
	\caption{Variation of the streamwise turbulence intensity, averaged over the rotor area, with distance, obtained using the original and optimized formulations of the wake-added turbulence, and compared with LES results from Ref. \cite{Eidi2022}. Subplots correspond to Cases (Opt.), (V-1), and (V-2), from top to bottom.}
	\label{fig:optAndVal}
\end{figure}

The distribution of disk-averaged streamwise turbulence intensity ($I_D$) behind the single-turbine of the optimization case is depicted in Figure~\ref{fig:optAndVal}(a). This figure clearly shows the overprediction of disk-averaged streamwise turbulence intensity through the original correlation of Crespo and Hern\'andez \cite{Crespo1996}. The local values of this quantity are consistently above those from LES which means that the recovery of the wake behind a turbine will be overestimated through the original correlation. On the other hand, focusing on the results obtained using the optimized formulation of wake-added turbulence, we see a good agreement with the high-fidelity data in the entire range, spanning from  $x/d_0 = 5$ to 50.

To further validate the local predictions of streamwise turbulence intensity behind a wind farm given by the optimized formulation, we apply it to Cases (V-1) and (V-2) and present the results in Figures~\ref{fig:optAndVal}(b) and (c), respectively. In this examination, we intentionally focus on the predictions $x/d_0 = 5$ after the last turbine, pursuing two key objectives. Initially, we aim to determine whether the new wake-added turbulence formulation, optimized based on a case with an incoming streamwise turbulence intensity of 7.25\% and combined with the non-cumulative superposition strategy, can accurately predict the evolution of streamwise turbulence intensity behind the last turbine in Cases (V-1) and (V-2). According to LES results, the last turbines in these cases experience higher incoming turbulence levels, around 15\% and 13\%, respectively. The results reveal, in both cases, that the new formulation satisfactorily predicts the disk-averaged streamwise turbulence intensity behind the last wind turbine with a good agreement with LES. The other objective of these validations is to observe how this quantity evolves further downstream in a farm. One can see a satisfactory match between the evolution of streamwise turbulence intensity predicted through the optimized wake-added turbulence formulation and LES, further downstream in both cases. This stands in contrast to the overestimation observed with the original formulation of the wake-added turbulence throughout the entire distance range.

\begin{figure}[!ht]
	\centering
    \includegraphics[width=0.6\textwidth]{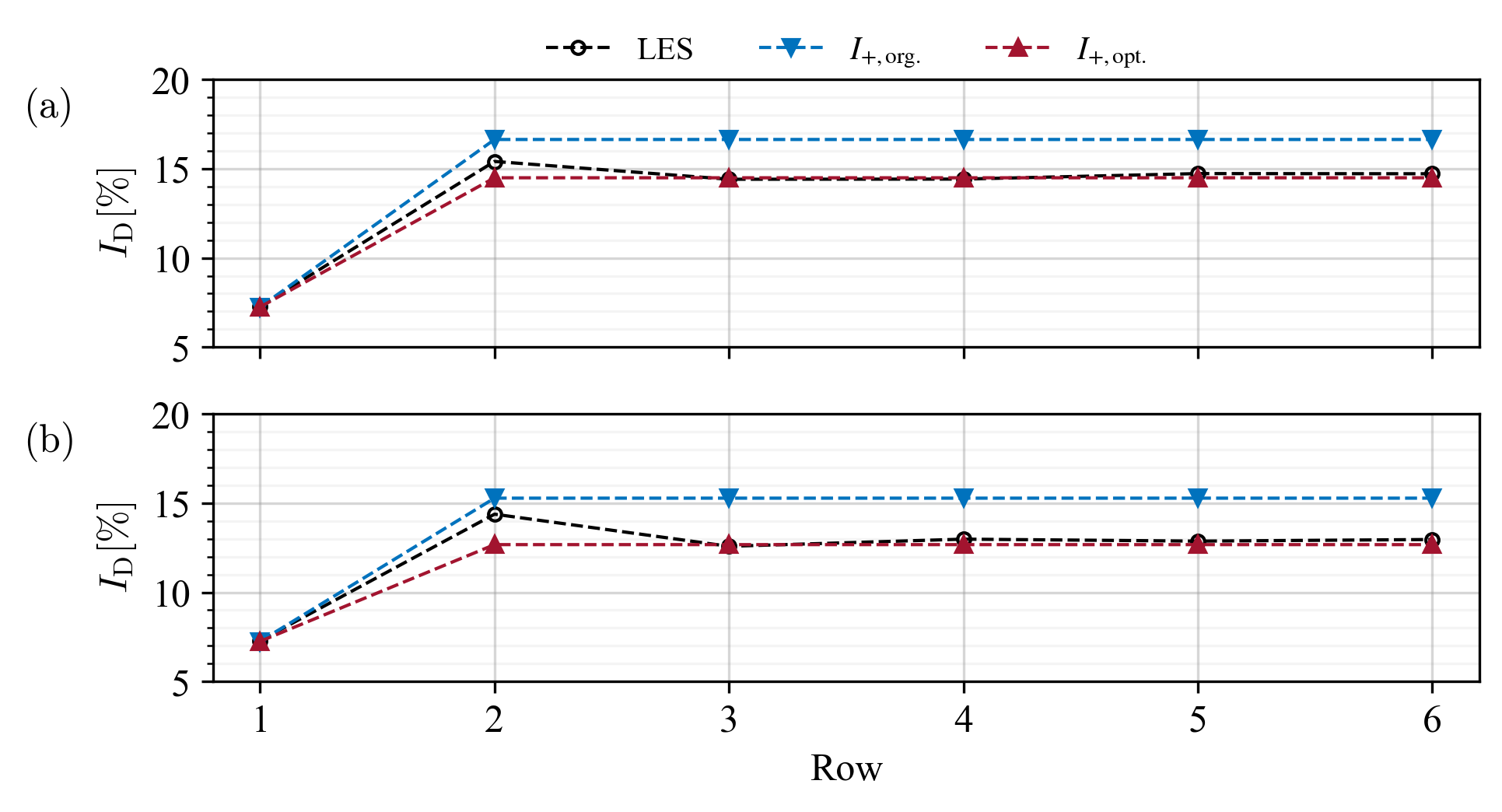}
	\caption{The disk-averaged streamwise turbulence intensity at the location of turbines, obtained using the original and optimized formulations of the wake-added turbulence, and compared with LES results from Ref. \cite{Eidi2022}. Subplots correspond to Cases (V-1), and Case (V-2), from top to bottom.}
	\label{fig:TI}
\end{figure}

To perform further quantitative validation, we focus on disk-averaged streamwise turbulence intensity at the location of turbines in Cases (V-1) and (V-2). This quantity will significantly impact the prediction of the wake growth rate and, consequently, serves as a crucial indicator for evaluating the optimized model's adequacy in predicting intra-farm wake losses. As shown in Figures~\ref{fig:TI}(a) and (b), respectively corresponding to Cases (V-1) and (V-2), we can infer that apart from the second row, the optimized model is consistently predicting closer-to-LES values of disk-averaged streamwise turbulence intensity at the location of turbines. 

\subsection{Intra-farm modeling }
To reveal the adequacy of the optimized wake-added turbulence model in predicting the intra-farm wake losses, here, we first combine it with GM and, afterward, use the framework to predict the normalized power of two $12 \times 6$ array, one with aligned and the other one with a staggered layout, i.e., Cases (IF-1) and (IF-2). Predictions will be compared against GM integrated with the original form of the wake-added turbulence model ($I_{+,\text{org.}}$), as well as the LES predictions from Ref. \cite{Stieren2022}. 

\begin{figure}[!ht]
	\centering
    \includegraphics[width=0.8\textwidth]{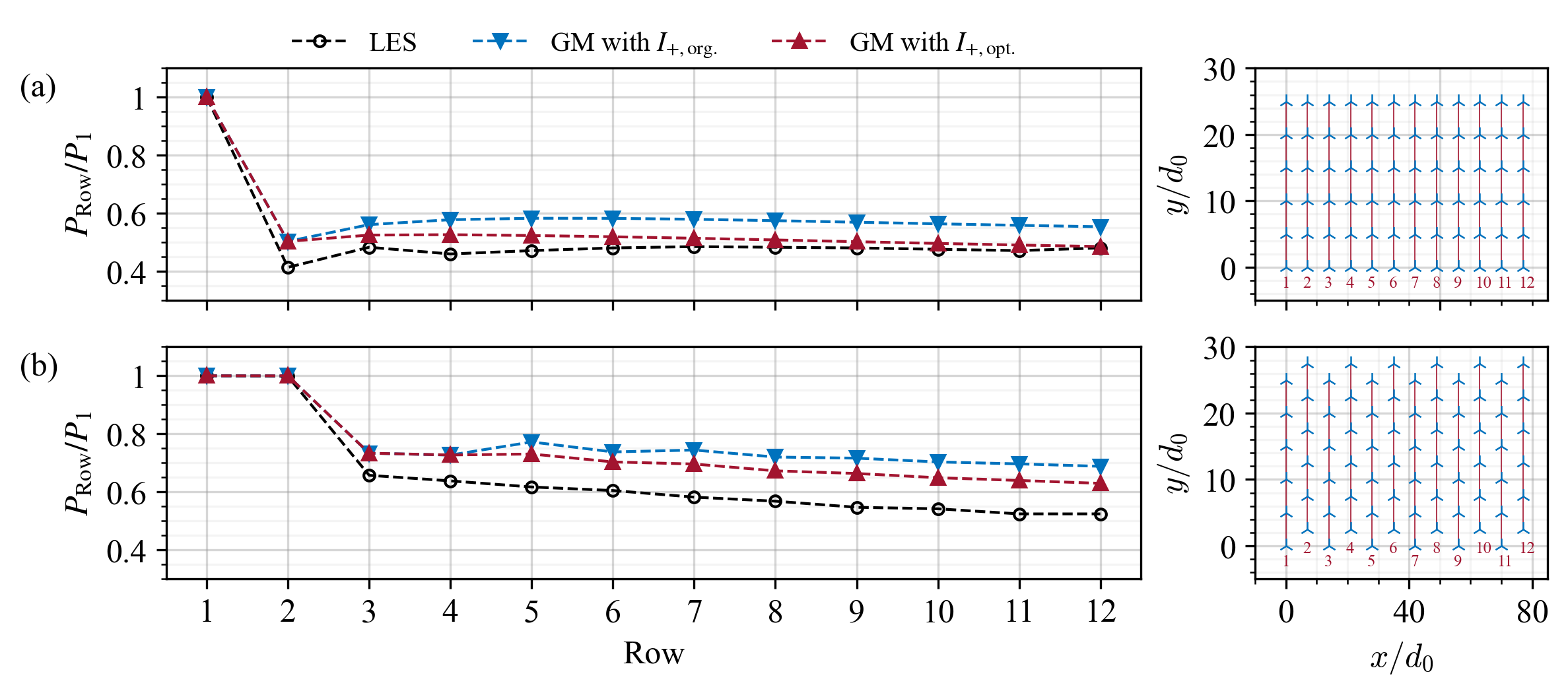}
	\caption{Left: The normalized power of turbine rows predicted by GM, integrated with original and optimized $I_{+}$ formulations, compared against LES outputs from Ref. \cite{Stieren2022}.  Right: The turbines marked by red lines are used to compute the average power for each row. The average power of the first row serves as the reference for calculating the normalized power. The subplots, from top to bottom, correspond to Cases (IF-1) and (IF-2), respectively. }
	\label{fig:NP_st_up}
\end{figure}

Figure~\ref{fig:NP_st_up}(a) illustrates the normalized power of turbine rows in Case (IF-1), where the streamwise spacing between the rows is $7d_0$. A comparison between GM integrated with $I_{+,\text{org.}}$ and LES results reveals a consistent overestimation across all turbine rows. This overestimation is attributed to the excessive prediction of wake recovery, which is linked to the high values of wake-added turbulence predicted by 
$I_{+,\text{org.}}$. In contrast, GM integrated with $I_{+,\text{opt.}}$ remedies this issue, consistently improving the predictions and shifting them closer to the LES outputs. 

Investigating the intra-farm wake in Case (IF-2), which features a staggered layout, reveals additional interesting findings, as shown in Figure~\ref{fig:NP_st_up}(b). The wake recovers more slowly in this case, which primarily accounts for the slight descending behavior observed in the LES predictions. This slow recovery can be partly attributed to the dense spanwise spacing of turbines in a staggered layout, which limits the expansion of wakes. Additionally, the deviation of wake growth from linear behavior, caused by the increased distance between consecutive turbines, may also contribute to this effect\cite{Stieren2021,Vahidi2022, Vahidi2022a, Souaiby2024}.

Table~\ref{tab:Table2} summarizes the performance of GM integrated with the optimized wake-added turbulence correlations in Cases (IF-1) and (IF-2). Here, the absolute errors between the wake models' predictions and LES results are calculated for each turbine row. The mean of these absolute error values across all turbine rows is then reported. We observe a marked improvement in predictions compared to those obtained using the original correlation of wake-added turbulence in both cases.
 
\begin{table}
\centering
\caption{A summary of the performance of GM, integrated with the original and optimized $I_{+}$ formulations, in predicting the normalized power when applied to Cases (IF-1) and (IF-2) and compared against LES outputs from Ref. \cite{Stieren2022}.}
\label{tab:Table2}
\setlength{\tabcolsep}{6pt}
\begin{tabular}{ccccccc}
\hline
& \multicolumn{2}{c}{Mean absolute relative error [\%]} & \\
Case   & GM with $I_{+,\text{org.}}$ & GM with $I_{+,\text{opt.}}$ & Error reduction [\%] \\ \hline
IF-1	&	18.10	&	7.40	&	10.70	\\
IF-2	&	21.04	&	15.14	&	5.90	\\
 \hline   
\end{tabular}
\end{table}

\begin{figure}[!ht]
	\centering
    \includegraphics[width=0.8\textwidth]{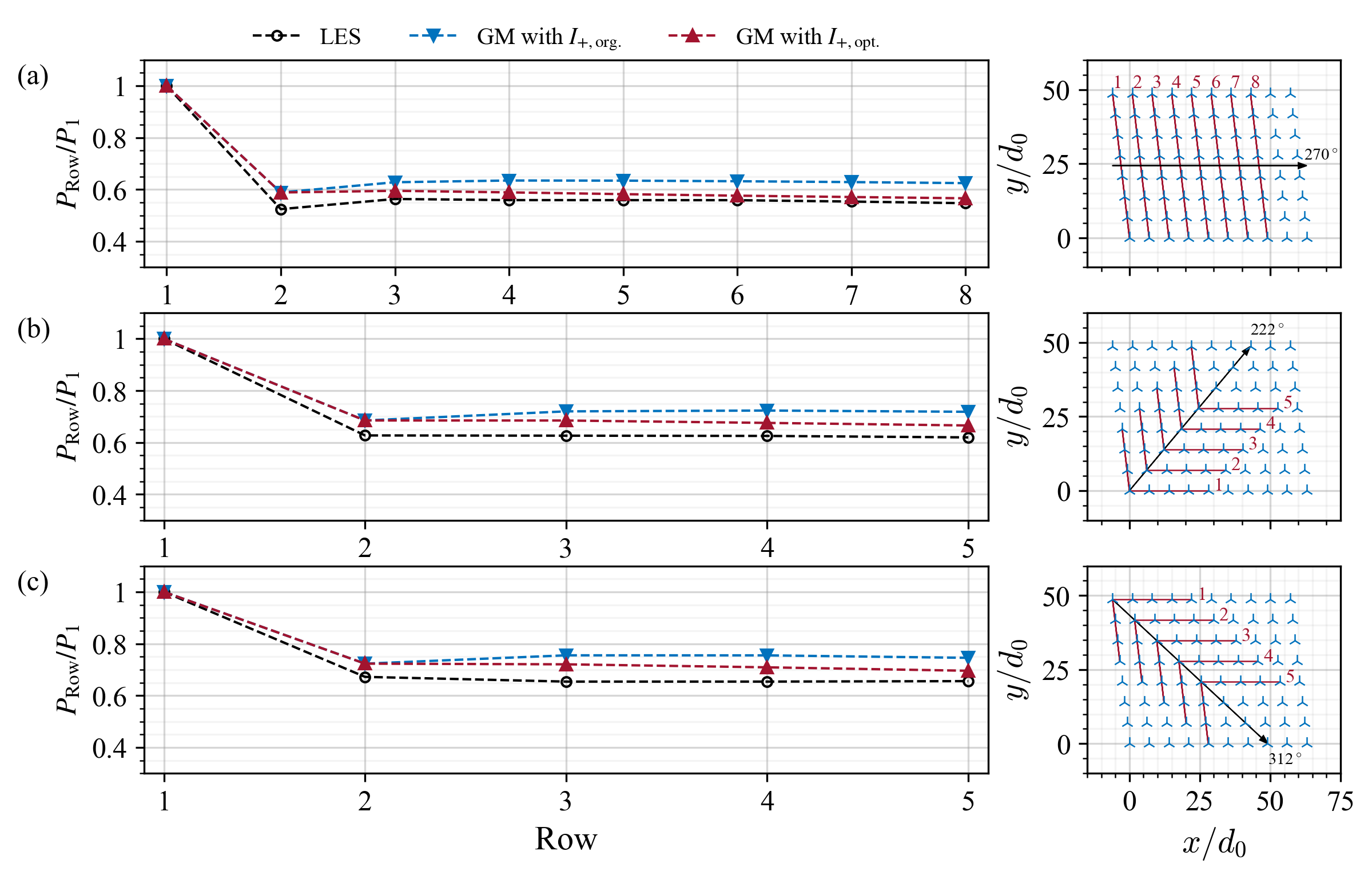}
	\caption{Left: The normalized power of turbine rows in Horns Rev 1 wind farm (Case (IF-3)) predicted by GM, integrated with the original and optimized $I_{+}$ formulations, compared against LES outputs from Ref. \cite{PorteAgel2013}. Right: The Horns Rev 1 wind farm's layout, with three characteristic wind directions. The turbines outlined in red are used to calculate the average power of rows. The first row's average power is the reference for normalizing the power values.}
	\label{fig:NP_HR1}
\end{figure}

To assess the improvements achieved in intra-farm modeling through the methodology proposed in this work, we apply the framework to a more realistic example based on the layout of the Horns Rev 1 wind farm. We study the normalized power of turbine rows in three wind directions of $270^\circ$, $222^\circ$, and $312^\circ$ in Figure~\ref{fig:NP_HR1}. The spacing between the consecutive turbines for these wind directions are $7d_0$, $9.3d_0$, and $10.4d_0$, respectively. It is evident that GM with $I_{+,\text{opt.}}$ consistently outperforms the model integrated with $I_{+,\text{org.}}$, and shows a better match with LES results reported in Ref. \cite{PorteAgel2013}. Table~\ref{tab:HR}, reports the performance of GM, integrated with the original and optimized $I_{+}$ formulations, in predicting the normalized power of turbine rows in Horns Rev 1 wind farm for these characteristic wind directions, showcasing the satisfactory performance of the GM model coupled with the optimized wake-added formulation.

\begin{table}
\centering
\caption{A summary of the performance of GM, integrated with the original and optimized $I_{+}$ formulations, in predicting the normalized power of turbine rows in Horns Rev 1 wind farm (Case (IF-3)), compared against LES outputs from Ref. \cite{PorteAgel2013}, in three incoming wind directions.}
\label{tab:HR}
\setlength{\tabcolsep}{6pt}
\begin{tabular}{ccccccc}
\hline
& \multicolumn{2}{c}{Mean absolute relative error [\%]} & \\
Wind direction [deg]   & GM with $I_{+,\text{org.}}$ & GM with $I_{+,\text{opt.}}$ & Error reduction [\%] \\ \hline
270	&	10.38	&	4.16	&	6.22	\\
222	&	9.95	&	6.07	&	3.88	\\
312	&	9.44	&	5.85	&	3.59	\\

 \hline   

\end{tabular}
\end{table}

\subsection{Farm-to-farm modeling}
\begin{figure}[!ht]
	\centering
    \includegraphics[width=0.6\textwidth]{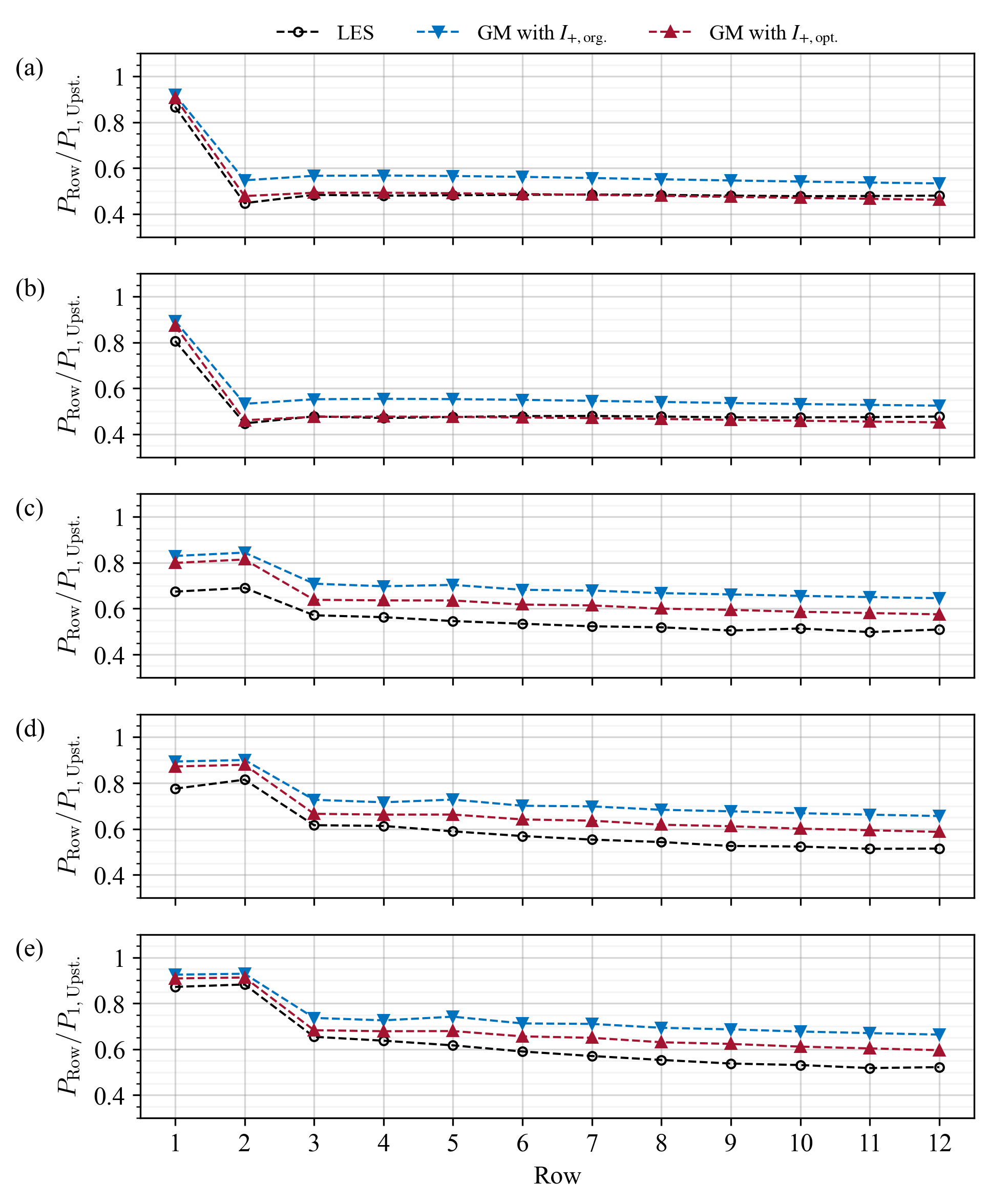}
	\caption{The normalized power of turbine rows of downstream wind farms predicted by GM, integrated with the original and optimized $I_{+}$ formulations, compared against LES outputs from Ref. \cite{Stieren2022}. Here, the average power of the first row of the upstream wind farm serves as the reference for calculating the normalized power. The subplots, from top to bottom, correspond to Cases (FF-1) to (FF-5).}
	\label{fig:NP_st_down}
\end{figure}

With the satisfactory impact of the optimized wake-added turbulence model on intra-farm wake modeling, it is time to apply it to farm-to-farm wake modeling. We first focus on Case (FF-1) where an aligned $12 \times 6$ array is positioned 10 km ($83.3d_0$) away from an aligned downstream counterpart. As shown in Figure~\ref{fig:NP_st_down}(a), integrating GM with $I_{+,\text{opt.}}$ improves the prediction of normalized power for the first row of the downstream wind farm. This shows that the optimized correlation for the wake-added turbulence could successfully address the issue of overestimated wind-farm wake recovery. Focusing on turbine rows 2 to 12 of the downstream wind farm, using $I_{+,\text{opt.}}$ consistently enhances the predictions, resulting in a good agreement with LES results. 

In Case (FF-2), the configuration is similar to the previous one while the upstream wind farm is a staggered array. As shown in Figure~\ref{fig:NP_st_down}(b), our predictions for the downstream turbines are affected by inaccuracies in the upstream predictions. The proposed methodology improves predictions for the first row of the downstream wind farm. Within the downstream wind farm, we also observe a good match between GM integrated with the optimized wake-added turbulence and LES results.

By studying Cases (FF-3) to (FF-5), we can investigate the role of the distance between the upstream and downstream wind farms, both of which have a staggered layout. As outlined earlier, in the staggered array considered in this work, the wake recovers slower and, thus, the wake of the upstream wind farm requires more space to recover in this scenario. We can see that the prediction of GM for the first row of the downstream turbine, regardless of the wake-added turbulence correlation, improves as the distance between the wind farms ($S_\text{WF}$) increases.

The performance of GM with both the original and optimized $I_{+}$ formulations in predicting the normalized power of turbine rows in the downstream wind farm for Cases (FF-1) to (FF-5) is compared against LES in Table~\ref{tab:Table4}. We observe that by integrating GM with $I_{+,\text{opt.}}$, the errors are reduced by approximately 10\% in all farm-to-farm investigations compared to the original formulation.

\begin{table}
\centering
\caption{A summary of the performance of GM, integrated with the original and optimized $I_{+}$ formulations, in predicting the normalized power when applied to Cases (FF-1) to (FF-5) and compared against LES outputs from Ref. \cite{Stieren2022}.}
\label{tab:Table4}
\setlength{\tabcolsep}{6pt}
\begin{tabular}{ccccccc}
\hline
& \multicolumn{2}{c}{Mean absolute relative error [\%]} & \\
Case   & GM with $I_{+,\text{org.}}$ & GM with $I_{+,\text{opt.}}$ & Error reduction [\%] \\ \hline
FF-1	&	14.72	&	2.40	&	12.32	\\
FF-2	&	14.02	&	2.76	&	11.26	\\
FF-3	&	27.06	&	15.68	&	11.38	\\
FF-4	&	22.70	&	12.64	&	10.06	\\
FF-5	&	20.05	&	10.79	&	9.26	\\
 
 \hline   

\end{tabular}
\end{table}

\section{Conclusions} \label{Sec:Conclusions}
This study aimed to develop an improved wake-added turbulence formulation to enhance the precision of intra-farm and farm-to-farm wake modeling through analytical frameworks. To accomplish this, we leveraged LES data from a single wind turbine and applied a grid-search optimization technique to refine the widely used wake-added turbulence correlation of Crespo and Hern\'andez\cite{Crespo1996}. Following this, we conducted comprehensive investigations into the evolution of streamwise turbulence intensity within and downstream of wind farms using two sets of in-house LESs. Finally, we evaluated the performance of the Gaussian wake model, integrated with both the original and optimized wake-added turbulence formulations, in modeling intra-farm and farm-to-farm wake effects using various LES datasets.

The results obtained from the intra-farm investigations, conducted with two $12 \times 6$ arrays, one aligned and one staggered, show that the new methodology can address the overestimation of wake recovery within wind farms. Additionally, the study of staggered layout indicated a slower recovery rate which could be linked to the concurrent increase in spacing between consecutive turbines and the decrease in spacing between turbine columns. Subsequently, studying Horns Rev 1 wind farm's efficiency in three incoming wind directions revealed that the Gaussian model coupled with the optimized wake-added turbulence consistently outperformed its counterpart with the original correlation. 

The analysis of farm-to-farm wake interactions in this study revealed that the Gaussian model, when used with the original wake-added turbulence formulation, consistently overestimates the normalized power of downstream wind farms. This overestimation is primarily due to an overprediction of wake recovery from the upstream wind farm. Across all farm-to-farm cases examined, integrating the Gaussian model with the optimized wake-added turbulence formulation led to improved predictions. Additionally, we explored the impact of the distance between upstream and downstream wind farms, both configured with staggered layouts. Adjusting the wake-added turbulence successfully resolved the overestimation of wake recovery in these cases as well.
\section*{CRediT authorship contribution statement}
\textbf{Navid Zehtabiyan-Rezaie}: Conceptualization, Methodology, Software, Formal analysis, Investigation, Data curation, Visualization, Writing – original draft. \textbf{Josephine Perto Justsen}: Methodology, Software, Formal analysis, Investigation. \textbf{Mahdi Abkar}: Conceptualization, Funding acquisition, Methodology, Project administration, Resources, Supervision, Writing – review \& editing. 
\section*{Conflict of interest}
The authors have no conflicts to disclose.
\section*{Data availability statement}
The data that support the findings of this study are available from the corresponding author upon reasonable request.
\section*{Acknowledgment}
The authors acknowledge the financial support from the Independent Research Fund Denmark (DFF) under Grant No. 0217-00038B. 
\bibliographystyle{elsarticle-num} 
\bibliography{Refs}





% \end{thebibliography}
\end{document}